\documentclass[prl, aps,
 twocolumn,
 superscriptaddress,
 amsmath,amssymb,
 10pt
]{revtex4-2}

\usepackage{times}
\usepackage{color}
\usepackage{graphicx}
\usepackage[dvipsnames]{xcolor}
\usepackage{physics}
\usepackage{bm}
\usepackage{mathtools}
\usepackage{upgreek}
\usepackage[colorlinks=true,allcolors=blue]{hyperref}
\usepackage{footmisc}
\usepackage{makecell}
\usepackage{mathtools}
\usepackage{soul}
\usepackage{xcolor}
\usepackage{lipsum}
\usepackage{wasysym}

\soulregister{\cite}{7}
\soulregister{\ref}{7}
\soulregister{\eqref}{7}
\soulregister{\label}{7}
\soulregister{\onlinecite}{7}
\definecolor{mycolor}{HTML}{ffffd4}
\sethlcolor{mycolor}

\begin{document}

\title{Disentangling the Impact of Quasiparticles and Two-Level Systems on the Statistics of Superconducting Qubit Lifetime}

\author{Shaojiang Zhu}
\email[szhu26@fnal.gov]{}
\affiliation{Superconducting Quantum Materials and Systems Center, Fermi National Accelerator Laboratory, Batavia, IL 60510, USA} 

\author{Xinyuan You}
\affiliation{Superconducting Quantum Materials and Systems Center, Fermi National Accelerator Laboratory, Batavia, IL 60510, USA}

\author{Ugur Alyanak}
\affiliation{Superconducting Quantum Materials and Systems Center, Fermi National Accelerator Laboratory, Batavia, IL 60510, USA}
\affiliation{Department of Physics, University of Chicago, Chicago, IL 60637, USA}

\author{Mustafa Bal}
\affiliation{Superconducting Quantum Materials and Systems Center, Fermi National Accelerator Laboratory, Batavia, IL 60510, USA}

\author{Francesco Crisa}
\affiliation{Superconducting Quantum Materials and Systems Center, Fermi National Accelerator Laboratory, Batavia, IL 60510, USA}

\author{\\Sabrina Garattoni}
\affiliation{Superconducting Quantum Materials and Systems Center, Fermi National Accelerator Laboratory, Batavia, IL 60510, USA}

\author{Andrei Lunin}
\affiliation{Superconducting Quantum Materials and Systems Center, Fermi National Accelerator Laboratory, Batavia, IL 60510, USA}

\author{Roman Pilipenko}
\affiliation{Superconducting Quantum Materials and Systems Center, Fermi National Accelerator Laboratory, Batavia, IL 60510, USA}

\author{Akshay Murthy}
\affiliation{Superconducting Quantum Materials and Systems Center, Fermi National Accelerator Laboratory, Batavia, IL 60510, USA}

\author{Alexander Romanenko}
\affiliation{Superconducting Quantum Materials and Systems Center, Fermi National Accelerator Laboratory, Batavia, IL 60510, USA}

\author{Anna Grassellino}
\affiliation{Superconducting Quantum Materials and Systems Center, Fermi National Accelerator Laboratory, Batavia, IL 60510, USA}

\begin{abstract}

Temporal fluctuations in the superconducting qubit lifetime, $T_1$, bring about additional challenges in building a fault-tolerant quantum computer.
Although exact mechanisms remain unclear, $T_1$ fluctuations are generally attributed to the strong coupling between the qubit and a few near-resonant two-level systems (TLSs) that can exchange energy with an assemble of thermally fluctuating two-level fluctuators (TLFs) at low frequencies. 
Here, we report $T_1$ measurements of qubits with different geometrical footprints and surface dielectrics as a function of temperature. 
By analyzing the noise spectrum of the qubit depolarization rate, $\Gamma_1 = 1/T_1$, we can disentangle the impact of TLSs, non-equilibrium quasiparticles (QPs), and equilibrium (thermally excited) QPs on the variance in $\Gamma_1$.
We find that $\Gamma_1$ variances in the qubit with a small footprint are more susceptible to the QP and TLS fluctuations than those in the large-footprint qubits.  
Furthermore, the QP-induced variances in all qubits are consistent with the theoretical framework of QP diffusion and fluctuation.
We suggest that these findings can offer valuable insight for future qubit design and engineering optimization.  

\end{abstract}

\maketitle

Superconducting qubits are a promising candidate for building a fault-tolerant quantum computer due to their flexibility in circuit design~\cite{gambetta2020ibm, arute2019quantum, blais2021circuit, gao2021practical, alam2022quantum}. This progress has been significantly propelled by the enhancement of the qubit lifetime, $T_1$, from the nanosecond to millisecond range over the past decades, advanced by improvements in material science, nanofabrication technology, and microwave design and engineering~\cite{siddiqi2021engineering, krinner2019engineering, krantz2019quantum}. Meanwhile, it has been observed that the lifetime $T_1$, as well as the decoherence time $T_2$ and the qubit frequency $\omega_\text{q}$, can fluctuate across the time and frequency domains; the qubits with higher $T_1$ show larger fluctuations~\cite{muller2015interacting, klimov2018fluctuations, schlor2019correlating, Burnett2019, bal2024systematic}. Since the gate fidelity is primarily limited by qubit coherence, these fluctuations will introduce additional challenges and obstacles to quantum error correction and future scalability.

Although the underlying physics is not yet fully understood, two-level defects residing in amorphous dielectrics at material surfaces are suggested as the primary source of decoherence in superconducting quantum devices~\cite{muller2019towards, carroll2022dynamics, wang2015surface, gao2008experimental, bilmes2020resolving}. High-frequency, near-resonance TLSs strongly couple to the qubit and dissipate energy to an environment at ambient temperature, causing qubit depolarization. Recently, M\"{u}ller \textit{et al.} pointed out that high-frequency TLSs can also interact with thermally fluctuating TLFs with $\hbar \omega \ll k_\text{B} T$
through dipole-dipole interactions, therefore fluctuating the qubit $T_1$~\cite{muller2015interacting}. Subsequent experimental studies of both flux-tunable and fixed-frequency transmon qubits support this theory by showing the signatures of a few TLSs responsible for the qubit lifetime fluctuation. Specifically, the spectral density of the qubit lifetime extracted from the measurements exhibit Lorentzian and $1/f$-type behavior~\cite{klimov2018fluctuations, Burnett2019, schlor2019correlating}.

Since the number of QPs can fluctuate due to mechanisms such as generation and recombination~\cite{DeVisser2011}, it is natural to hypothesize that these exact mechanisms can lead to fluctuations in the coherence of superconducting qubits. 
To elucidate the observed power law relationship between the average and standard deviation of the qubit $T_1$, the assumption that the qubit $T_1$ is limited by the population of thermally excited QPs was made for this purpose~\cite{simbierowicz2021qubit, li2023power}. 
The same assumption was used to explain the temporal fluctuations of the qubit $T_1$ extracted from the double exponential decay in the relaxation curves~\cite{yan2016flux}.
By examining the $T_1$ fluctuations at a few selected temperatures, M\"{u}ller \textit{et al.} briefly discussed the contributions of both TLS and QPs to the $T_1$ fluctuations in the same qubit and suggested that if there is any, the fluctuation due to QPs is likely to be smaller than the one associated with strongly coupled TLSs~\cite{muller2015interacting}. 

Although many studies have focused on understanding the qubit $T_1$ fluctuations induced by TLS or QP, it is interesting to investigate and disentangle both fluctuating sources in the same device to better understand the decoherence mechanisms and propose strategies for mitigating these noise channels.
In this work, we perform $T_1$ measurements on three fixed-frequency transmon qubits with different geometrical footprints and surface dielectrics over a temperature range of 7-153 mK. 
Continued $T_1$ measurements of $\sim$72 hours at each temperature allow us to analyze the spectral density of qubit fluctuations in terms of the depolarization rate, $\Gamma_1$, across a broad range of frequencies, from which we can extract the contributions of the TLS and QPs to the variance in $\Gamma_1$.
We further analyze the QP densities near the qubit junction based on a diffusion model and provide quantitative support on the experimental observations of qubit $\Gamma_1$ variances induced by non-equilibrium and equilibrium QPs, respectively.

We label the three qubits under study as A, B, and C, all fabricated on two separate but identical Sapphire chips (A and B are on one chip and C on another). Both chips share the same qubit layout, and each qubit has its own $\lambda / 4$ coplanar waveguide resonator that is inductively coupled to a common feedline for qubit dispersive readout. Qubits A and B use a 160 nm Nb thin film for the patterned capacitor pads, while qubit C has additionally \textit{in situ} sputtered $\sim$10 nm Ta encapsulated on the top of the Nb film.
Qubit A has small pads (120$\times$510 $\mu \text{m}^2$) separated by a 20 $\mu$m gap, and qubits B and C adopt the same geometry, specifically, 150$\times$720 $\mu \text{m}^2$ pads with a 150 $\mu$m gap~\cite{Supplementary_Zhu}. The fabrication processes are kept nominally identical for the two chips, and the exact size of Al/AlO$_x$/Al junction is used for all qubits. The two chips are separately packaged in the same type of enclosures (Cu coated with Au), wired on two separate measurement chains with nominally identical configurations, and characterized in the same cooldown. 
Additional details on qubit fabrication, material characterization, and microwave measurements can be found in Ref.~\onlinecite{bal2024systematic}. Regarding TLS losses, our microwave simulation based on a 3D model shows that the energy participation ratio (EPR) of the surface dielectric in qubit A is about two times larger than in qubits B and C~\cite{Supplementary_Zhu}.
The dielectric losses of the surface oxide on metal pads of qubits A and B are the same, reported to be on the order of $\delta_{\text{NbO}_x} \sim 10^{-2}$~\cite{romanenko2020three, zhu2022high}; the dielectric loss $\delta_{\text{TaO}_x} \sim 10^{-3}$ \cite{crowley2023disentangling} of qubit C is about an order of magnitude lower than that of qubits A and B.

\begin{figure}
   \centering
   \includegraphics[width=\columnwidth]{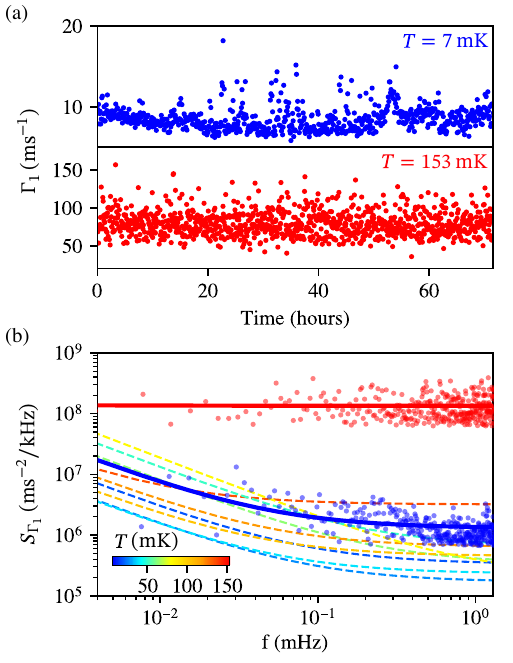}
   \caption{(a) Typical $\Gamma_1$ temporal fluctuations of qubit B at 7 mK (blue) and 153 mK (red), respectively. At low temperatures, typical telegraphic noise of $\Gamma_1$ fluctuations with instantaneous switching and finite dwell time exhibits signatures of the TLS-TLF interaction. At high temperatures, the white noise behavior of $\Gamma_1$ indicates that the QP fluctuations are the dominating noise source. (b) The noise spectrum of qubit depolarization rate $S_{\Gamma_1}$ at 7 and 153 mK are plotted as the solid symbols. The two data sets are fitted with equation $S_{\Gamma_1}(f) = a/f + b$ and shown as the solid lines. The dashed lines are the fitted curves at all other temperatures, illustrating the transition of the dominating noise source from TLSs at low temperatures to QPs at high temperatures.
   }
   \label{fig:fig_1} 
\end{figure}

At each elevated temperature, we wait two hours to thermalize the qubits before performing qubit $T_1$ measurements consecutively for about 72 hours. The measurements of qubits A and B are interleaved, while qubit C's are continuous.
Each $\Gamma_1$ point in Fig.~\ref{fig:fig_1} is extracted by fitting a single-exponential decay of the qubit energy-relaxation curve that typically takes 6 to 10 minutes, depending on the lifetime length of each qubit.
In the following analyses, we assume that the TLSs and QPs primarily determine the depolarization rate of our qubits due to the following facts: (1) the Purcell rate of the qubit is experimentally verified to be below 1 kHz, (2) the loss due to trapped flux in the pads is negligible because all qubits are fixed-frequency transmons with a single, sub-micron Josephson junction, and (3) the enclosure is designed to have the lowest resonance mode at 10 GHz, well above the qubit frequency (approximately 4 GHz). 
Therefore, we can write the qubit depolarization rate as the sum of the TLS and QP losses, \textit{i.e.}, $\Gamma_1 = \Gamma_\text{TLS} + \Gamma_\text{QP}$, with $\Gamma_\text{TLS} = p \delta_\text{TLS}$. Here $p$ is the geometry-dependent EPR, and $\delta_\text{TLS}$ is determined by the surface dielectric of the metal pads. In this work, we only consider the dielectric loss from the metal-air interface~\cite{bal2024systematic}.

Figure~\ref{fig:fig_1}(a) shows the temporal $\Gamma_1$ fluctuations of qubit B at 7 mK (blue) and 153 mK (red), respectively. 
At low temperatures, the $\Gamma_1$ fluctuation resembles similar features as reported in ~\cite{muller2015interacting, klimov2018fluctuations, schlor2019correlating, Burnett2019}, exhibiting signatures of the TLS-TLF interaction.
At higher temperatures, the $\Gamma_1$ fluctuation shows a white noise behavior, consistent with the behavior resulting from fluctuations in the QP number~\cite{DeVisser2011}.
To disentangle the contributions of TLSs and QPs to the $\Gamma_1$ fluctuations, we convert the temporal fluctuations to spectral density by performing the Fourier transform of the autocorrelation function of the qubit depolarization rate $\langle \Gamma_1(t) \Gamma_1(0) \rangle$, following the method reported in Ref.~\onlinecite{muller2015interacting}. The obtained spectral densities of qubit B at 7 mK (blue dot) and 153 mK (red dot) are shown in Fig.~\ref{fig:fig_1}(b) as a function of frequency. 
At low temperatures (\textit{e.g.}, $T=7$ mK), we observe a clear $1/f$ behavior at low frequency and a white noise behavior at high frequency. The former behavior is predicted within the interacting TLS-TLF model, where a few TLSs undergo frequency fluctuations due to interactions with an assembly of thermal TLFs~\cite{muller2015interacting}. On the other hand, it is known that the number of QPs can fluctuate due to the dynamics of QP generation and recombination. This results in a Lorentzian-type spectral density~\cite{wilson2001time, DeVisser2011}. Since the frequencies we measured here are much smaller than the characteristic QP recombination rate ($\sim$ kHz), the Lorentzian profile can be well approximated as constant white noise. Therefore, we fit the data with the equation $S_{\Gamma_1}(f) = a/f + b$, with results shown as thick solid lines in Fig.~\ref{fig:fig_1}(b). 
At higher temperatures (\textit{e.g.}, $T=153$ mK), the spectral density exhibits mainly white noise behavior, indicating that QPs dominate the fluctuation. In Fig.~\ref{fig:fig_1}(b), we also plot the fitted curves at different temperatures to illustrate the transition of the dominant noise source from TLSs at low temperatures to QPs at high temperatures.

\begin{figure}[t]
\centering
\includegraphics[width=\columnwidth]{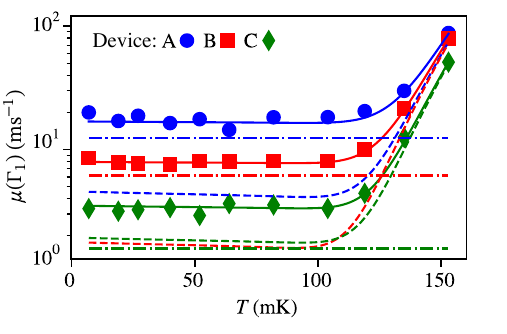}
\caption{\label{fig:fig_2} Temperature dependence of the average depolarization rate \(\mu(\Gamma_1)\) for all three qubits. Solid symbols represent the data, dash-dotted lines are the fitted contribution from TLS, and dashed lines are the fitted contribution from non-equilibrium and equilibrium QPs. Solid lines show the sum of the TLS and QP losses, which is in agreement with the measured data. The extracted fitting parameters are provided in Table~\ref{tab:example}.}
\end{figure}

We next analyze the impact of QP and TLS losses on the qubit average $\Gamma_1$, $\mu(\Gamma_1)$, as a function of temperature. The measurement data is shown as solid symbols in Fig.~\ref{fig:fig_2}. It is seen that $\mu(\Gamma_1)$ of all three qubits show weak temperature dependence when temperatures are below $\sim$100 mK, then exponentially increase and approach the same value around 153 mK. 
Based on established TLS and QP theories, we can unravel the contributions of TLS and QP loss to $\mu(\Gamma_1)$. First, although both the relaxation and excitation rate of the qubit due to a single TLS depend on temperature, the sum of both, \textit{i.e.}, the total TLS depolarization rate contributing to the qubit is temperature independent~\cite{You2021, You2022}.
Secondly, the contributions of QPs to the qubit $\Gamma_1$ can be written as~\cite{DeVisser2011, serniak2018hot, Glazman2020}:
\begin{equation}
    \Gamma_{\text{QP}} = \frac{16 E_\text{J}}{\hbar\pi}\sqrt{\frac{E_\text{C}}{8E_\text{J}}}\sqrt{\frac{2\Delta}{\pi k_\text{B}T}}\exp\left(\frac{\hbar\omega_\text{q}}{2k_\text{B}T}\right)K_0\left(\frac{\hbar\omega_\text{q}}{2k_\text{B}T}\right)x_\text{QP},
\end{equation}
where \( E_\text{J} \) and \( E_\text{C} \) are the Josephson and charging energies of the qubit with transition frequency \( \omega_\text{q} \). \( K_0(z) \) is a modified Bessel function. \( x_\text{QP} \) is defined as the QP number density \( n_\text{QP} = N_\text{QP}/V \) normalized by the Cooper pair number density \( n_\text{CP}=2\nu_0\Delta \), with $\nu_0\approx 1(\text{eV}\text{\r{A}}^3)^{-1}$~\cite{Glazman2020}. The temperature dependence of \( x_\text{QP} \) is modeled as \( x_\text{QP} = x_\text{QP}^0 + \sqrt{2\pi k_\text{B}T/\Delta}\exp(-\Delta/k_\text{B}T) \), where the first term describes the non-equilibrium QP density (\(x^0_\text{QP}\)) that is temperature independent, and the second term denotes the equilibrium (thermal) QP density (\(x^\text{th}_\text{QP}\)) that is exponentially dependent on the temperature. 
Using experimentally determined \( E_\text{J} \), \( E_\text{C} \), and \( \omega_\text{q} \), we can fit the data to extract the TLS loss rate \( \Gamma_\text{TLS} \), the non-equilibrium QP density \( x_\text{QP}^0 \), and the superconducting gap \( \Delta \). Fitted results are shown in Table~\ref{tab:example}~\footnote{In the fitting, we assume the $x_\text{QP}^0$ to be the same for devices B and C, as they share the same pad and junction design.}.

From the fits, all three qubits show similar gap energy \(\Delta/2\pi \sim 40\) GHz, which is consistent with the values reported in the literature~\cite{connolly2024coexistence, kamenov2023suppression}, although slightly lower. This is likely due to differences in Al film thicknesses and fabrication processes. 
The extracted \(\Gamma_\text{TLS}\) values are shown as dash-dotted lines in Fig.~\ref{fig:fig_2}, and the improvement of $\Gamma_\text{TLS}$ from qubit A to C can be explained by either the simulated EPR or the differences in dielectric loss.
The ratio of \(\Gamma_\text{TLS}\) between qubits A and B is approximately 1.9, which matches our EPR simulation; the ratio between qubit B and C is about 3.9, smaller than the reported \(\delta_{\text{NbO}_x} / \delta_{\text{TaO}_x} \sim 10\)~\cite{crowley2023disentangling, zhu2022high, romanenko2020three}. This discrepancy might arise because the Ta thin film in qubit C encapsulates only the top surface of Nb but not the sidewall~\cite{bal2024systematic}. 
The contribution from QPs is shown as dashed lines in Fig.~\ref{fig:fig_2}.
As expected, \(\Gamma_\text{QP}\) is dominated by \(x^\text{th}_\text{QP}\) at temperatures, $k_\text{B} T$, above the Al gap, $\Delta$, and by \(x^0_\text{QP}\) when $k_\text{B} T$ is below $\Delta$.
However, it should be noted that although the junctions in all three qubits are nominally identical, the fitted \(x^0_\text{QP}\) shows geometric dependence, with \(x^0_\text{QP}\) ratio $\sim 2.5$ between small pads (qubit A) and large pads (qubits B and C). 
We prove that this difference can be qualitatively explained in the following, with more details in the Supplementary Material~\cite{Supplementary_Zhu}.

We adopt a phenomenological energy-dependent diffusion model~\cite{Riwar_Quasi, Martinis_Quasi, alyanak2024modeling} to calculate the QP density arriving at the junction, assuming that the non-equilibrium QPs generated by high-energy radiation distribute uniformly in the qubit pads and elastically diffuse to the junction. The incoming QP energy is set to be near the gap energy of the system, and Neumann boundary conditions in the substrate-pad interface and junction are flux continuity.
In the non-equilibrium regime, $x^0_{\text{QP}}$ in qubit A is about $2.7$ times larger than the ones in qubits B and C due to the shorter distance that QPs travel in small pads to the junction. 
This value is close to the ratio obtained from experimental fitting.
In the equilibrium regime, QPs are dominated by those thermally generated from the low-gap Al (instead of high-gap Nb). Since all qubits studied in this experiment share the same design of Al/AlO$_x$/Al Josephson junction, we expect $x^{\text{th}}_{\text{QP}}$ to be the same.

\begin{figure}[t]
\centering
\includegraphics[width=\columnwidth]{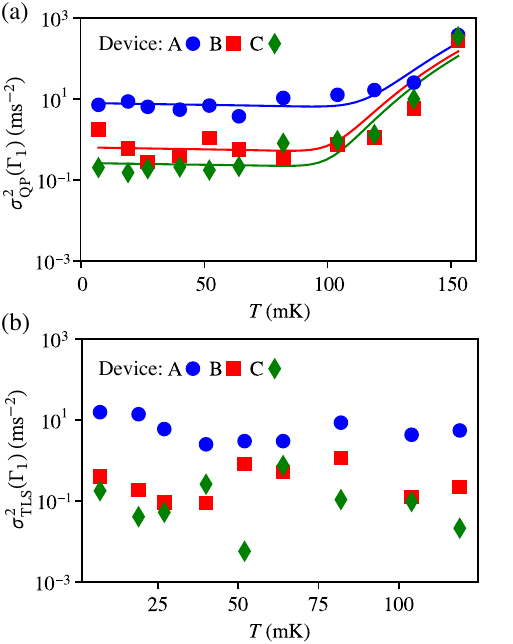}
\caption{\label{fig:fig_3} Temperature dependence of qubit \(\Gamma_1\) variance contributed by (a) QPs and (b) TLSs. In Figure (a), solid symbols represent the data extracted from experiments, and solid lines are the fitting based on Eq.~\eqref{eq:2}. The extracted fitting parameters are provided in Table~\ref{tab:example}.}
\end{figure}

To quantify the contributions of QPs and TLSs to the variance of the qubit relaxation rate \(\Gamma_1\), we integrate the white noise and \(1/f\) noise spectral densities over the measured frequency range at each temperature [see the fitted curves shown in Fig.~\ref{fig:fig_1}(b)], which allows us to separately determine the variances of the qubit depolarization rates induced by QPs, \(\sigma^2_{\text{QP}}(\Gamma_1)\), and TLSs, \(\sigma^2_{\text{TLS}}(\Gamma_1)\). The results are plotted as solid symbols in Fig.~\ref{fig:fig_3}(a) and (b).

We first discuss the impact of QP fluctuations on the $\Gamma_1$ variance. 
From Fig.~\ref{fig:fig_3}(a), we can see that $\sigma^2_\text{QP}(\Gamma_1)$ in all qubits exhibit a very similar trend: a weak temperature-dependence at \(T<100\) mK followed by a rapid increase at higher temperatures.
It is also seen that in the low-temperature regime, qubit A is much more susceptible to the QP fluctuations, with $\sigma^2_\text{QP}(\Gamma_1)$ at least an order of magnitude higher than those observed for qubits B and C, while the latter two have similar $\Gamma_1$ variances. With the QP diffusion and fluctuation model, we can provide quantitative support in understanding these observations and further disentangle the impacts of non-equilibrium and thermal QPs on the qubit fluctuations.
Assuming the number of QPs follows a Poisson distribution~\cite{DeVisser2011}, we have the QP variance \(\sigma^2(N_\text{QP})\) equals its mean value \(\mu(N_\text{QP})\).
Therefore, the variance \(\sigma^2_\text{QP}(\Gamma_1)\) is given by~\cite{Supplementary_Zhu}:
\begin{equation}\label{eq:2}
    \sigma^2_{\text{QP}}(\Gamma_1) = \eta^2(T)\left( \frac{x_\text{QP}^0}{n_\text{CP}V^0_\text{eff}} + \frac{x_\text{QP}^\text{th}}{n_\text{CP}V^\text{th}_\text{eff}} \right),
\end{equation}
with \(\eta(T)=\frac{16 E_\text{J}}{\hbar\pi}\sqrt{\frac{E_\text{C}}{8E_\text{J}}}\sqrt{\frac{2\Delta}{\pi k_\text{B}T}}\exp\left(\frac{\hbar\omega_\text{q}}{2k_\text{B}T}\right)K_0\left(\frac{\hbar\omega_\text{q}}{2k_\text{B}T}\right)\).
Here, the non-equilibrium and equilibrium QPs are assumed to be independent~\cite{serniak2018hot}. In Eq.~\eqref{eq:2}, we define \(V^\text{0}_\text{eff}\) and \(V^\text{th}_\text{eff}\) as the effective volumes for non-equilibrium and thermal QPs, respectively, to reflect their distinct contributions to the fluctuations. Using previously obtained $x^0_{\text{QP}}$, $x^{\text{th}}_{\text{QP}}$, and $\Delta$ [see Table~\ref{tab:example}], we can fit the experimentally extracted $\sigma^2_\text{QP}(\Gamma_1)$ and obtain the values of \(V^\text{0}_\text{eff}\) and \(V^\text{th}_\text{eff}\) for each qubit. 
The fitted curves are shown as solid lines in Fig.~\ref{fig:fig_3}(a), and the fitted \(V^\text{0}_\text{eff}\) and \(V^\text{th}_\text{eff}\) are also listed in Table~\ref{tab:example}.
In the non-equilibrium regime, the effective volume is expected to be proportional to $(v_\text{F} \cdot \tau_\text{r})^3$, with the Fermi velocity $v_\text{F}$ being the same for all qubits, and $\tau_\text{r}$ the QP recombination rate that is inversely proportional to \(x_\text{QP}^0\)~\cite{DeVisser2011}. 
Therefore, we have \(V^0_\text{eff} \propto (1/x_\text{QP}^0)^3\).
We notice that the ratio of \(V^0_\text{eff}\) between device C (B) and A is about 13 (5) from the fitting.
Indeed, this aligns with our data: the ratio of \(x_\text{QP}^0\) between device A and C (B) is about 2.5 (\(2.5^3 \sim 15\)). 
Moreover, the difference in \(x_\text{QP}^0\) can be supported by the diffusion process, from which the \(x_\text{QP}^0\) ratio is calculated to be 2.7 between different pad sizes. 
In the thermal regime, the QP density dominated by the thermal generation in the Al junction leads increases significantly but does not substantially suppress the \( V_\text{eff}^\text{th} \). 
On the other hand, \( V_\text{eff}^\text{th} \) cannot exceed the upper limit set by the non-equilibrium QPs from the qubit pads when \(k_\text{B} T\) approaches the Al gap energy.
Therefore we approximate the \( V_\text{eff}^\text{th} \) as a constant in Eq.~\eqref{eq:2}. From the fit, we find the effective volume $V^\text{th}_\text{eff}$ for all the qubits approaches the same value, close to the volume of the Josephson junction \(\sim 0.013 \, \mu \textrm{m}^3\), which is as expected since all qubits consist of the same Al junctions. Here, we define the junction volume as the product of the junction area and the total thickness of Al layers.

\begin{table}[t]
\caption{\label{tab:example} Fitting parameters extracted for three different transmon qubits. Here, \(x_\text{QP}^0\) is the density of non-equilibrium QPs, \(\Delta\) is the superconducting gap energy of Al, \(V_\text{eff}^0\) and \(V_\text{eff}^\text{th}\) are the effective volumes for non-equilibrium and thermal QPs, and \(\Gamma_\text{TLS}\) is the depolarization rate from TLS.}
\begin{ruledtabular}
\begin{tabular}{cccccc}
{Qubit} & {$x_\text{QP}^0$} & {$\Delta/2\pi$ (GHz)} & {$V_\text{eff}^0$ ($\mu\text{m}^3$)} & {$V_\text{eff}^\text{th}$ ($\mu\text{m}^3$)} & {$\Gamma_\text{TLS}$ ($\mu\text{s}^{-1}$)}  \\ 
\\[-0.3cm]
\hline
\\[-0.3cm]
A & $1.4\times 10^{-7}$  & 38.0 & 0.062 & 0.025 & 1.2$\times 10^{-2}$ \\ 
B & $5.5\times 10^{-8}$  & 38.2 & 0.290 & 0.039 & 6.2$\times 10^{-3}$\\ 
C & $5.5\times 10^{-8}$  & 39.6 & 0.807 & 0.037 & 1.6$\times 10^{-3}$ \\ 
\end{tabular}
\end{ruledtabular}
\end{table}

In Fig.~\ref{fig:fig_3}(b), we plot the $\sigma^2_{\text{TLS}}(\Gamma_1)$ induced by TLSs. Compared to the variance from QPs, the $\sigma^2_{\text{TLS}}(\Gamma_1)$ here show an overall weak dependence on temperature, with some local features. 
It is also observed that $\sigma^2_{\text{TLS}}(\Gamma_1)$ of qubit A is notably larger than qubits B and C, while the latter two show some consistency in the temperature range.
We suggest that this can be qualitatively explained by TLS theory as well.
The variance of qubit depolarization due to the interaction between a single TLS and an assemble of TLFs is given by~\cite{You2022}:
\begin{equation}\label{eq:3}
    \sigma^2_\text{TLS}(\Gamma_1) = A^4 \frac{16\gamma_2^2\omega_\delta^2}{(\gamma_2^2 + \omega_\delta^2)^4}\left[ \sum_i g_i^2 \left(1-\tanh^2\frac{\omega_\text{t,i}}{2k_\text{B}T}\right)  \right]^2,
\end{equation}
where \(A\) denotes a constant related to the coupling strength and qubit matrix element, \(\gamma_2\) describes the linewidth of the TLS, and \(g_i\) the coupling strength between the TLS and the \(i\)th TLF, with frequency \(\omega_\text{t,i}\). 
When the temperature increases, the TLS linewidth gets larger, suppressing the amplitude of the Lorentzian and, thus, the variance of the fluctuation. On the other hand, higher temperature leads to more frequent switching of the TLFs, increasing the last factor in Eq.~\eqref{eq:3}. This discussion is based on a single TLS. In practice, multiple TLSs can couple simultaneously to the qubit, each with different dephasing rates and frequencies. As a result, the above two competing factors for different TLSs may lead to the observed behavior in Fig.~\ref{fig:fig_3}(b)~\cite{Supplementary_Zhu}.

Our work underscores a practical need to stabilize qubit lifetimes. On one hand, theoretical proposals based on driving the qubits~\cite{Matityahu2021} or the TLSs~\cite{You2022} could provide ways to stabilize the fluctuations due to TLSs without affecting the average lifetime. However, protocols to reduce QP fluctuations have yet to be developed, but may benefit from strategies to reduce QPs in general~\cite{Martinis_Quasi, iaia2022phonon, riwar2016normal,kamenov2023suppression}.

We perform prolonged, statistical $T_1$ measurements on three superconducting transmon qubits across a broad temperature range. Different geometrical footprints and superconducting materials of these qubits enable us to disentangle the impacts of QPs and TLSs on the average and fluctuation of the qubit lifetime.
The improvements in the average qubit lifetime can be explained by an optimization of the qubit design as well as a reduced loss of surface dielectric. More importantly, the QP densities near the Josephson junction depend on the qubit footprints, \textit{i.e.}, higher density in the small-footprint qubit. 
We further propose the concept of effective QP volume, which scales cubically with the inverse of the non-equilibrium QP density and provides quantitative support for understanding the discrepancy between the lifetime fluctuations in qubits with different footprints. 
We suggest that these findings can guide future qubit design and engineering optimization.

\begin{acknowledgments}
This material is based upon work supported by the U.S. Department of Energy, Office of Science, National Quantum Information Science Research Centers, Superconducting Quantum Materials and Systems Center (SQMS) under contract number DE-AC02-07CH11359.
\end{acknowledgments}

\bibliography{main.bib}

\end{document}


\title{Supplementary Material for 
``Disentangling the Impact of Quasiparticles and Two-Level Systems on the Statistics of Superconducting Qubit Lifetime''}

\author{Shaojiang Zhu}
\affiliation{Superconducting Quantum Materials and Systems Center, Fermi National Accelerator Laboratory, Batavia, IL 60510, USA} 

\author{Xinyuan You}
\affiliation{Superconducting Quantum Materials and Systems Center, Fermi National Accelerator Laboratory, Batavia, IL 60510, USA}

\author{Ugur Alyanak}
\affiliation{Superconducting Quantum Materials and Systems Center, Fermi National Accelerator Laboratory, Batavia, IL 60510, USA}
\affiliation{Department of Physics, University of Chicago, Chicago, IL 60637, USA}

\author{Mustafa Bal}
\affiliation{Superconducting Quantum Materials and Systems Center, Fermi National Accelerator Laboratory, Batavia, IL 60510, USA}

\author{Francesco Crisa}
\affiliation{Superconducting Quantum Materials and Systems Center, Fermi National Accelerator Laboratory, Batavia, IL 60510, USA}

\author{\\Sabrina Garattoni}
\affiliation{Superconducting Quantum Materials and Systems Center, Fermi National Accelerator Laboratory, Batavia, IL 60510, USA}

\author{Andrei Lunin}
\affiliation{Superconducting Quantum Materials and Systems Center, Fermi National Accelerator Laboratory, Batavia, IL 60510, USA}

\author{Roman Pilipenko}
\affiliation{Superconducting Quantum Materials and Systems Center, Fermi National Accelerator Laboratory, Batavia, IL 60510, USA}

\author{Akshay Murthy}
\affiliation{Superconducting Quantum Materials and Systems Center, Fermi National Accelerator Laboratory, Batavia, IL 60510, USA}

\author{Alexander Romanenko}
\affiliation{Superconducting Quantum Materials and Systems Center, Fermi National Accelerator Laboratory, Batavia, IL 60510, USA}

\author{Anna Grassellino}
\affiliation{Superconducting Quantum Materials and Systems Center, Fermi National Accelerator Laboratory, Batavia, IL 60510, USA}

\maketitle

\section{Qubit layout}

\begin{figure}[b]
\centering
\includegraphics[width=.4\columnwidth]{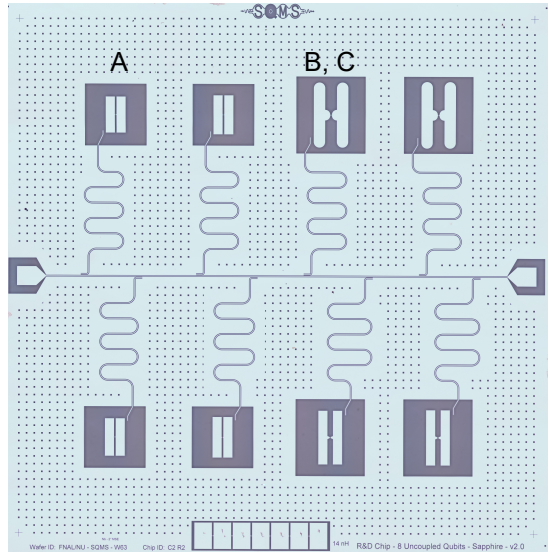}
\caption{\label{fig:fig1s} Microscope image of the qubit chip, showing all eight isolated qubits with each own $\lambda/4$ CPW resonator that inductively coupled to the common feedline. The capacitor pad size of qubit A is $120 \times 510$ $\mu\text{m}^2$ with 20 $\mu\text{m}$ gap, and the one of qubit B and C are $150 \times 720$ $\mu\text{m}^2$ with 150 $\mu\text{m}$ gap.}
\end{figure}

\begin{figure}[t]
\centering
\includegraphics[width=.5\columnwidth]{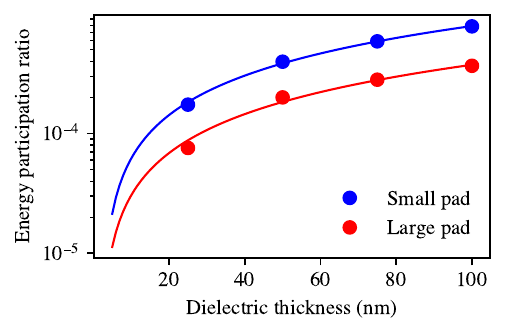}
\caption{\label{fig:fig2s} Simulated EPR for small pad (blue dot) and large pad (red dot). Solid curves are the linear fit of each data set, from which we can extrapolate the EPR at the thickness of 5 nm for both pad sizes. The EPR ratio of the two pads at 5 nm is about 2.}
\end{figure}

Figure~\ref{fig:fig1s} shows the qubit chip layout. The shunt capacitors of the eight fixed-frequency transmon qubits on the Sapphire chip with 7.5 $\times$ 7.5 mm$^2$ size are grouped into three geometrical footprints with different surface participation ratios for coherence comparison. Each qubit is capacitively coupled to an independent $\lambda/4$ CPW resonator; each resonator is inductively coupled to a common feedline through which the microwave signal is transmitted. With the same qubit layout design, we performed a systematical study on the qubit lifetime $T_1$ by encapsulating different materials on top of the same Nb base layer. All fabrication processes, Josephson junctions, and measurement configurations and methods are kept nominally identical~\cite{bal2024systematic}. 

The three qubits under discussion in this work are labeled as A, B, and C on the layout. The shunt capacitors of Qubit A and B on one chip are made with a single Nb base layer; the ones of qubit C on another chip have an encapsulation of a thin ($\sim 10 nm$) Ta layer on top of the Nb base. All qubits are designed to operate around 4 GHz, and the resonances of readout resonators are between 7-8 GHz.

\section{Energy participation ratio simulation}
We perform the energy participation ratio (EPR) simulation with a 3D microwave model, which is necessary for the qubit layout due to its irregular geometries compared with the CPW resonator~\cite{wenner2011surface, woods2019determining}. 

Because of the extraordinary geometrical aspect ratio of the qubit pads dimensions (hundreds of micro-meters) to the thickness of the thin dielectric layer (a few nano-meters) on the surface of the metal pads, careful considerations of the tradeoff between simulation accuracy and cost are taken into account. Unlike the other strategies in the full 3D qubit simulations~\cite{wang2015surface, gambetta2016investigating}, we first adopt thick ($200$ nm) dielectric to perform the simulation with large meshing size. We gradually reduce the dielectric thickness down to $20$ nm without sacrificing the simulation accuracy (meshing size) to get the EPR values at different dielectric thicknesses. We find a good linear relationship between EPR and dielectric thickness, therefore we can obtain the EPR at $5$ nm by extrapolating the simulated data (see Figure ~\ref{fig:fig2s}). During the simulation, we also carefully considered the actual profiles of the etched film, such as a certain degree of the film sidewall slope, the rounded edge corner, and the very shallow trenches on the Sapphire substrate due to over-etching. Based on the material characterizations, we only consider the metal-air interface in the simulation, using the measured dielectric thickness of $5$ nm with dielectric constant $\epsilon = 10$~\cite{wenner2011surface, wang2015surface, woods2019determining}.

We find that the EPR of qubit A is about two times larger than qubit B's. We also find that, for qubit C, the EPR of the surface (TaO$_x$) and the sidewall (NbO$_x$) are approximately the same, so the effective dielectric loss combining both materials is about 5.5 times smaller than the loss of NbO$_x$, which can qualitatively explain the $\sim 4$ times improvement of TLS loss from qubit B to C in the experiments.

\section{Temperature Dependence of Qubit Lifetime Induced by Two-Level Systems}

In our analysis of qubit lifetime as a function of temperature, we assume the contribution from two-level systems (TLSs) to be independent of temperature. Below, we present the derivation supporting this assumption.
According to the standard TLS model~\cite{muller2019towards}, the noise spectral density, \( S_{\text{TLS}}(\omega) \), of a single TLS is given by:
\begin{equation}
    S_{\text{TLS}}(\omega) = \frac{1 + \langle \sigma_z \rangle_{\text{eq}}}{2} \frac{2\gamma_2}{(\omega - \omega_0)^2 + \gamma_2^2} + \frac{1 - \langle \sigma_z \rangle_{\text{eq}}}{2} \frac{2\gamma_2}{(\omega + \omega_0)^2 + \gamma_2^2},
\end{equation}
where \( \langle \sigma_z \rangle_{\text{eq}} = \tanh{\left(\frac{\omega_0}{2k_{\text{B}}T}\right)} \) represents the equilibrium polarization, dependent on temperature \( T \), \( \gamma_2 \) is the TLS dephasing rate, and \( \omega_0 \) is the resonance frequency of the TLS. When the TLS is coupled to a qubit, the first term corresponds to a Lorentzian centered at positive frequency \( \omega_0 \) and primarily contributes to qubit relaxation, whereas the second term, centered at negative frequency \( -\omega_0 \), predominantly results in qubit excitation.

The same temperature dependence gives rise to the phenomenon that TLSs can be saturated at high temperatures to enhance the resonator Q, which is inversely proportional to the difference of the excitation and relaxation process, \textit{i.e.,} $\langle \sigma_z \rangle_\text{eq}$.
However, the previous statement is no longer valid in the context of qubit lifetime. Contrary to being dictated by the difference between the relaxation and excitation processes, the qubit's depolarization rate (inverse of lifetime) is governed by their summation.
Therefore, the dependence on $\langle \sigma_z \rangle_\text{eq}$ is suppressed (by considering a qubit is in close resonance with the TLS).
Moreover, the TLS dephasing rate \( \gamma_2 \) may exhibit a modest temperature dependence. However, an ensemble of TLSs with a specific distribution renders the averaged spectral density effectively temperature-independent~\cite{You2021}.

\section{Quasiparticle density calculation}
To calculate the ratio of $x^0_{\text{QP}}$ that reach the junction for different geometries, we use a 1D phenomenological diffusion model \cite{Riwar_Quasi, Martinis_Quasi} that assumes Neumann boundary conditions in the pad-substrate interface and the Josephson Junction. This choice makes sure the current will be continuous both during phonon-mediated quasiparticle creation at the substrate and quasiparticle tunneling across the junction. We use the following diffusion equation,
\begin{align}
\frac{\partial \rho_\epsilon(x,t)}{\partial t}
   = D(\epsilon,x)\frac{\partial^2 \rho_\epsilon(x,t)}{\partial x^2}- \frac{\rho_\epsilon(x,t)}{\tau(\epsilon,x)}+j(\epsilon, x,t),
\end{align}
where $\rho_\epsilon(x,t)$ is the spatio-temporal quasiparticle density, $\tau$ is the QP lifetime, and $D$ is the diffusion coefficient. 
In this 1D model, we set the length of the system to be $L$, the distance between the substrate-pad interface and the junction. The incoming quasiparticle energy is assumed to be around $\Delta$, and $j(\epsilon, x,t)$ is the injection term non-zero only for $t=0$ and a small neighborhood of $x=0$. With this information, we can set up a relationship between $x^{0}_{\text{QP}}$ and $\rho_\epsilon(x,t)$:
\begin{align}
x^{0}_{\text{QP}}\equiv \int_{0}^{t_{\text{max}}}\mathrm{d}t'\,\rho_{(\epsilon\approx\Delta)}(L,t'),
\end{align}
where $t_{\text{max}}$ is set to be much longer than the expected QP time of flight from the pad to the junction and shorter than the time the system takes to generate another QP bunch due to external sources. Since the measurable quantity that affects $T_1$ and its fluctuations is the density that tunnels across the junction, and incoming energy is near $\Delta$, we set $x=L$ and $\epsilon=\Delta$.

\begin{figure}[t]
\centering
\includegraphics[width=.5\columnwidth]{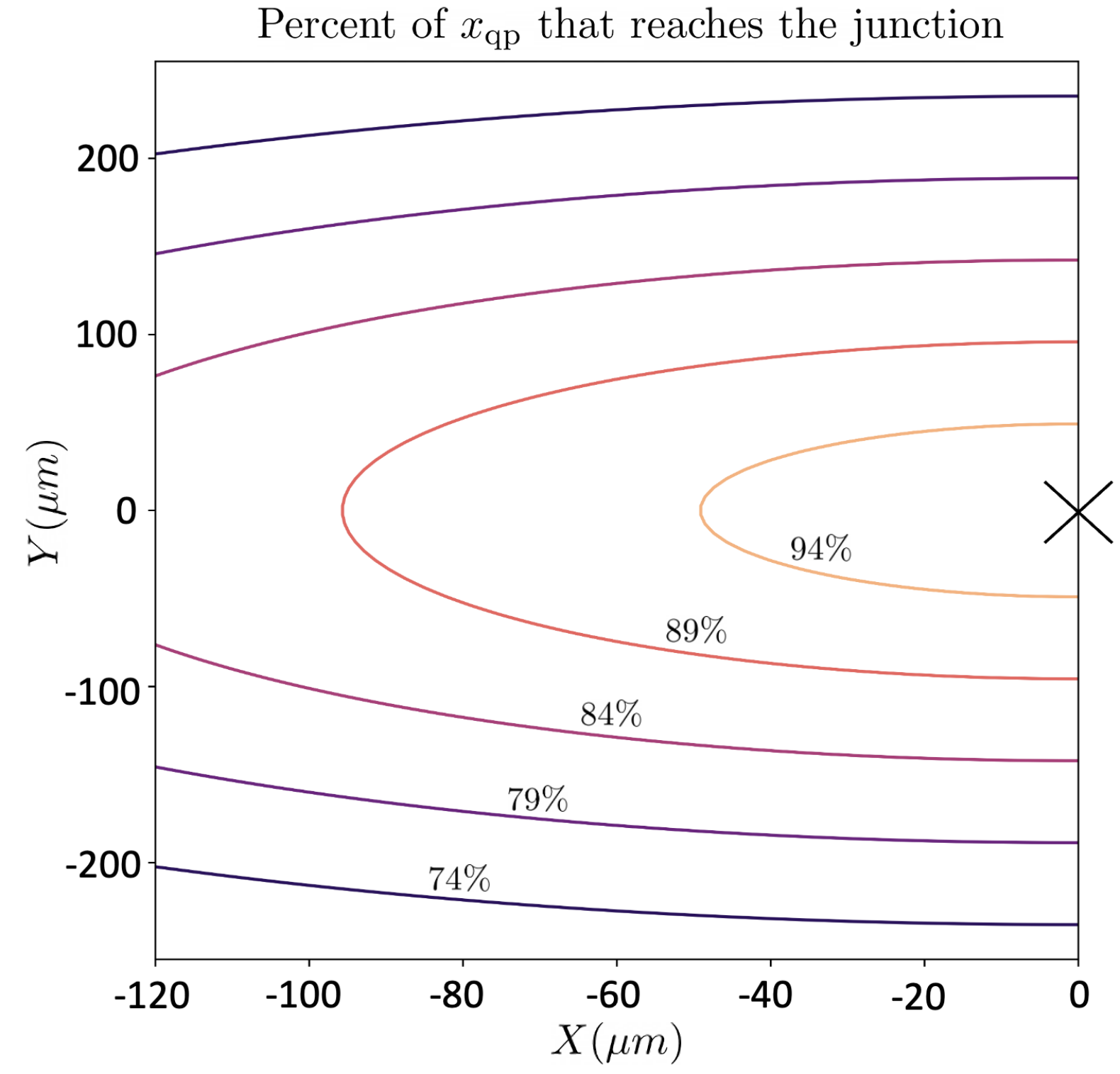}
\caption{\label{fig:fig4s} Percent of non-equilibrium quasiparticle density $x^{0}_{\text{QP}}$ that reaches the junction given incoming position $(X, Y)$ calculated for the small pad geometry (qubit A), where the junction lies at $(0,0)$ indicated with a cross. }
\end{figure}

The remaining step is to convert our 3D junction geometry to a 1D length. Instead of finding the mean distance between the pad and junction, we calculate the percentage of density reached at the junction using our 1D model for all possible paths the QP can take across the pad geometry and average these percentages for small (qubit A) and large (qubit B and C) pad geometries (see Figure~\ref{fig:fig4s} for percentages of qubit A). The results show that assuming the incoming QP densities are the same for all pads, the $x^{0}_{\text{QP}}$ reaching the junction between pad A and pads B (C) has a factor of 2.73 difference. 

\section{Derivation of Eq. (2) in the Main Text}
In this section, we derive Eq. (2) as presented in the main text. Starting from Eq. (1), the quasiparticle (QP) decay rate is given by
\begin{equation}
\Gamma_{\text{QP}} = \eta(T) x_\text{QP},
\end{equation}
where \( \eta(T) \) is defined as
\begin{equation}
\eta(T) = \frac{16 E_\text{J}}{\hbar \pi} \sqrt{\frac{E_\text{C}}{8 E_\text{J}}} \sqrt{\frac{2 \Delta}{\pi k_\text{B} T}} \exp\left(\frac{\hbar \omega_\text{q}}{2 k_\text{B} T}\right) K_0\left(\frac{\hbar \omega_\text{q}}{2 k_\text{B} T}\right),
\end{equation}
with \( E_\text{J} \) and \( E_\text{C} \) being the Josephson and charging energies, respectively, and \( K_0 \) denoting the modified Bessel function of the second kind.
To determine the variance of \( \Gamma_{\text{QP}} \), we consider the fluctuations in the QP density \( x_\text{QP} \), yielding
\begin{equation}
    \sigma^2_{\text{QP}}(\Gamma_1) = \eta^2 \sigma^2(x_\text{QP}).
\end{equation}
Expressing \( x_\text{QP} \) in terms of the QP number \( N_\text{QP} \) and volume \( V \), we find
\begin{equation}
\sigma^2(x_\text{QP}) = \sigma^2\left(\frac{N_\text{QP}}{V}\right) = \frac{1}{V^2} \sigma^2(N_\text{QP}).
\end{equation}
Assuming that \( N_\text{QP} \) follows a Poisson distribution, we have \( \sigma^2(N_\text{QP}) = N_\text{QP} \), leading to
\begin{equation}
\sigma^2_{\text{QP}}(\Gamma_1) = \eta^2 \frac{x_\text{QP}}{V} .
\end{equation}
Finally, we assume that thermal and non-equilibrium QPs are independent of each other, allowing us to write
\begin{equation}
    \sigma^2_{\text{QP}}(\Gamma_1) = \eta^2 \left( \frac{x_\text{QP}^0}{n_\text{CP} V_\text{eff}^0} + \frac{x_\text{QP}^\text{th}}{n_\text{CP} V_\text{eff}^\text{th}} \right),
\end{equation}
where the superscripts “0” and “th” denote non-equilibrium and thermal QPs, respectively.

\begin{table}[t]
\caption{\label{tab:example} Parameters used in Eq. (3) from the main text for the simulation of the two-TLS model shown in Fig.~\ref{fig:fig5s}. }
\begin{ruledtabular}
\begin{tabular}{cccccc}
{TLS} & 
{$A$ (a.u.)} & 
{$\gamma_2$ (MHz)} & 
{$\omega_\delta/2\pi$ (MHz)} & 
{$g_{i=1,2,3}/2\pi$ (MHz)} & 
{$\omega_\text{ti=1,2,3}/2\pi$ (MHz)}  \\ 
\\[-0.3cm]
\hline
\\[-0.3cm]
1 & 2.1  &  $T/(40 \text{mK})$ & 2.0 & [10, 10, 10] & [100, 200, 300] \\ 
2 & 1.0  & $8T/(40 \text{mK})$ & 1.0 & [10, 10, 10] & [100, 200, 300]\\ 
\end{tabular}
\end{ruledtabular}
\end{table}

\begin{figure}[b]
\centering
\includegraphics[width=.5\columnwidth]{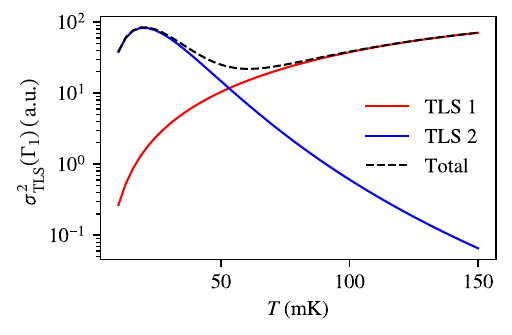}
\caption{\label{fig:fig5s} 
Temperature dependence of $\sigma^2_\text{TLS}(\Gamma_1)$ obtained from two TLS model. Parameters used in the simulation are shown in Table~\ref{tab:example}. }
\end{figure}

\section{Temperature dependence of $T_1$ fluctuation due to TLSs}

In Fig. 3 from the main text, we observe that the fluctuation from TLS only $\sigma^2_\text{TLS}(\Gamma_1)$ exhibits weak temperature dependence and displays some local minimum. 

We provide possible explanations that this might be due to two competing effects: 
(1) Higher temperature leads to more frequent switching of the TLFs, increasing the fluctuation variance.
(2) Higher temperature results in larger TLS linewidth, suppressing the amplitude of the noise spectrum and, thus, the variance of the fluctuation. 
Here, we perform a simple numerical simulation to support the above statement. 
Specifically, we consider two TLSs to be near-resonant with the qubits. 
Table~\ref{tab:example} shows all the parameters used in the simulation [see Eq. (3) and parameter definition thereafter in the main text]. 
Specifically, we take the dephasing rate $\gamma_2$ of both TLSs on the order of MHz~\cite{Lisenfeld2016-jr}, with a linear temperature dependence~\cite{PhysRevX.13.041005}. 
In Fig.~\ref{fig:fig5s}, we plot the contribution from two TLSs and their sum. 
Notably, each TLS's contribution exhibits marked temperature dependence, with TLS1 initially exhibiting an increase, followed by a rapid decay, while TLS2 continues to increase steadily across the temperature range. However, the combined contributions of the TLSs demonstrate a weaker overall temperature dependence, characterized by the presence of a local minimum that closely resembles the behavior reported in the main text.

\bibliography{main.bib}